\documentstyle[psfig,aps,prl,multicol]{revtex}
\begin{document}
\title{Entropic Bell Inequalities}
\author{N. J. Cerf$^1$ and C. Adami$^{1,2}$}
\address{$^1$W. K. Kellogg Radiation Laboratory and 
         $^2$Computation and Neural Systems\\
California Institute of Technology,
Pasadena, California 91125, USA}

\date{29 August 1996}

\draft
\maketitle
\vskip -0.1cm
\begin{abstract}
We derive entropic Bell inequalities from considering entropy Venn
diagrams. These entropic inequalities, akin to the Braunstein-Caves
inequalities, are violated for a quantum mechanical
Einstein-Podolsky-Rosen pair, which implies that the
conditional entropies of Bell variables must be negative in this
case.  This suggests that the satisfaction of entropic Bell
inequalities is equivalent to the non-negativity of conditional entropies
as a necessary condition for separability.
\end{abstract}
\pacs{PACS numbers: 03.65.Bz,05.30.-d,89.70.+c
      \hfill KRL preprint MAP-205}
\vspace{-0.4cm}
\begin{multicols}{2}[]
\narrowtext

The essence of Bell inequalities\cite{bib_bell,bib_general} is related
to Einstein's notion of ``realism''~\cite{bib_EPR}: that an object has
``objective properties'' whether they are measured or not. Bell
inequalities, in their simplest form, reflect constraints on the
statistics of any {\em three} local properties of a collection of
objects. These constraints must be obeyed if the three properties can
be independently known for each object. 
An intuitive discussion of Bell inequalities in
this context is due to Wigner
\cite{bib_wigner,bib_despagnat}. Consider a set of objects, each
characterized by three two-valued (or dichotomic) properties $a$, $b$,
and $c$. Then, grouping the objects as a function of {\em two} (out of
the three) properties (for instance grouping together objects having
property $a$ but not $b$), it is easy to build a simple inequality
relating the number of objects in various groups defined by different
pairs of properties. For example,
\begin{equation}   \label{eq_wigner}
n(a,{\rm not~}b) \le n(a, {\rm not~}c) + n({\rm not~}b,c)\;.
\end{equation}
While such an inequality only refers to the simultaneous specification
of any {\em pair} of properties, its satisfaction depends on the
existence of a probability distribution for all {\em three}.  Thus,
even when the three properties can not be accessed at the same time
(for whatever reason), Eq. (\ref{eq_wigner}) still holds provided that
there exists such an {\it objective} description of each object using
three parameters $a$, $b$, and $c$; therefore, Eq.~(\ref{eq_wigner})
provides a straightforward test of ``local realism'' (i.e., the
combination of objectivity and locality).  
As confirmed experimentally~\cite{bib_aspect},
an inequality such as Eq.~(\ref{eq_wigner}) can be violated
in quantum mechanics. It is the uncertainty
principle (implying that the simultaneous perfect
knowledge of two conjugate observables is impossible) which is at the
root of such a violation.
Arguments similar to those above are used to
derive the Bell inequalities\cite{bib_bell} or their generalization,
the Clauser-Horne-Shimony-Holt (CHSH) inequalities\cite{bib_CHSH}, and
their violation can be traced back to the nonexistence of an
underlying joint probability distribution involving incompatible
variables.

\par

The purpose of this paper is to show that the violation of Bell
inequalities in quantum mechanics is directly connected to the
existence of {\it negative} quantum entropies, a
feature which is classically forbidden.  We have shown
in previous work\cite{bib_neginfo} that a consistent quantum
information theory treating classical correlation and quantum
entanglement on the same footing implies that conditional entropies
can be negative. This purely quantum behavior can be traced back to
the fact that the eigenvalues of a ``conditional density matrix'' can
exceed one. (In contrast, the conditional probabilities in classical
information theory are always bounded by one, which implies the
classical property that conditional entropies are non-negative.)
Negative conditional entropies appear precisely in the case of quantum
entanglement~\cite{bib_neginfo}, for instance for an
Einstein-Podolsky-Rosen (EPR) wavefunction, which is the typical
object of Bell-type experiments.  As a consequence, it is natural to
seek for a relation between this non-classical feature and the
violation of Bell inequalities, the standard evidence for the
existence of quantum nonlocal correlations.  To begin with, we derive
an {\em entropic} Bell inequality that resembles the conventional 
one but involves mutual entropies rather than correlation
coefficients. This entropic Bell inequality is related to the
Braunstein-Caves information Bell inequality\cite{bib_braunstein} 
and implies Schumacher's triangle inequality for information
distances\cite{bib_schumacher}. Unlike those, however, our inequality
has a structure isomorphic to the conventional one, and has a simple
geometric interpretation based on the ternary entropy diagram
describing the Bell variables $a$, $b$, and $c$.  Indeed, we show that the
violation of our entropic Bell inequalities implies that one
out of three conditional entropies describing $abc$ must be negative,
a feature that eliminates any classical description of the system. We
show that these entropic Bell inequalities are violated when
performing Bell-type measurements on EPR pairs, for example, but not
necessarily at the same angles as the conventional Bell
inequalities. Therefore, our entropic Bell inequalities
provide another necessary condition for separability, distinct from
the standard Bell inequalities.  

\par
Consider two widely separated entangled systems in general, or, more
specifically, a pair of spin-1/2 particles in a singlet 
state (Bohm's~\cite{bib_bohm} version of an EPR pair)
\begin{equation}
|\Psi\rangle = {1 \over \sqrt{2}}\Big( |\uparrow \downarrow \rangle
                                 -|\downarrow \uparrow \rangle\Big)\;.
\end {equation}
Assume an observer, acting independently on each particle, can measure
the spin component of that particle along two possible orientations,
for example with a Stern-Gerlach setup.  Let the first observer either
measure the $z$-component of one of the particles (and call this
observable $A$ and the outcome of the measurement $a$) or else the
component along an axis making an angle $\theta$ with the $z$-axis
(observable $B$, with outcome $b$).  Correspondingly, the second
observer measures (on the second particle) either the $z$-component
(observable $A'$) or else the component making an angle $\phi$ with
the $z$-axis (observable $C$)~\cite{fn1}.  Locality implies that the two
distant observers have no influence on each other, i.e., the decision
to make one of the two possible measurements on the first particle
does not affect the outcome of the measurement on the other
particle. Indeed, it is known that the marginal statistics of the
outcome of the spin measurement on the second particle, $c$ for
instance, is unchanged whether one measures $A$ or $B$ on the first
particle. Let us now outline a general derivation of conventional Bell
inequalities (see, e.g.,~\cite{bib_peres}).  Consider 3 dichotomic
random variables $A$, $B$, and $C$ that represent properties of the
system and can only take on the values +1 or $-1$ with
equal probability (1/2). For our purposes,
they stand of course for the measured spin components (either up or
down along the chosen axis), i.e., the Bell variables.  (As $A'$ is
fully anticorrelated with $A$, we do not make use of it.)  Any random set
of outcomes $a$, $b$, and $c$ must obey
\begin{equation}  \label{eq_peres}
ab+ac-bc \le 1
\end{equation}
along with the two corresponding equations obtained by
cyclic permutation ($a \to b \to c$). Indeed, 
the left-hand-side of Eq.~(\ref{eq_peres}) is equal to 1 when $a=b$, while
it is equal to $-1\pm 2$ when $a=-b$. Taking the average of
Eq.~(\ref{eq_peres}) and its permutations yields the three basic 
Bell inequalities
\begin{eqnarray}
\label{eq_bell1}
\langle ab \rangle + \langle ac \rangle - \langle bc \rangle &\le& 1\\
\label{eq_bell2}
\langle ab \rangle - \langle ac \rangle + \langle bc \rangle &\le& 1\\
\label{eq_bell3}
-\langle ab \rangle + \langle ac \rangle + \langle bc \rangle &\le& 1
\end{eqnarray}
relating the correlation coefficients between pairs of
variables.  Eqs.~(\ref{eq_bell2}) and (\ref{eq_bell3}) can
be summarized in the form of the standard Bell
inequality~\cite{bib_peres}
\begin{equation}  \label{eq_bell}
| \langle ab \rangle - \langle ac \rangle | + \langle bc \rangle\le 1\;. 
\end{equation}
The important point is that inequalities
(\ref{eq_bell1}-\ref{eq_bell3}) involve only the simultaneous
specification of {\em two} (out of the three) random variables,
although it is assumed that the {\em three} variables possess an
element of reality, i.e., they can {\em in principle} be known at the
same time, even if not in practice. In other words, it is assumed that
there exists an underlying joint probability distribution $p(a,b,c)$,
in which case the Bell inequalities (which depend only on the marginal
probability $p(a,b)=\sum_c p(a,b,c)$ and cyclic permutations) must be
satisfied.  Therefore, the violation of {\em any} of the inequalities
(\ref{eq_bell1}-\ref{eq_bell3}) implies that $a$, $b$, and $c$ cannot
derive from a joint distribution (i.e., cannot be described by any
local hidden-variable theory), as emphasized in
Ref.~\cite{bib_braunstein}.  In the following, we will show that
the violation of Bell inequalities, while ruling out such a classical
underlying description of local realism, still does not contradict a
quantum one based on an underlying joint density matrix $\rho_{ABC}$,
but forces the corresponding entropies to be {\em
negative}~\cite{bib_neginfo}.

\par
\begin{figure}[t]
\caption{Ternary entropy diagram for the Bell variables $ABC$. The
entries $\alpha,\bar\alpha,\beta,\bar\beta,\gamma,\bar\gamma$ and
$\delta$ are defined in the text. All of them (except $\delta$) are
non-negative in Shannon information theory~\protect\cite{bib_ash}.}
\vskip 0.25cm
\centerline{\psfig{figure=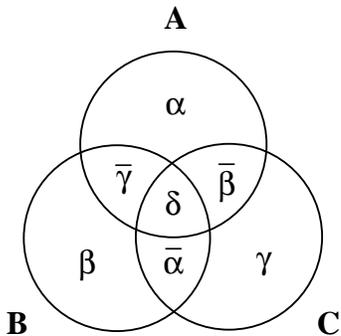,width=1.75in,angle=-90}}
\label{fig_ternary}
\vskip -0.25cm
\end{figure}

Let us derive Bell inequalities akin to the
conventional ones, Eqs.~(\ref{eq_bell1}-\ref{eq_bell3}), but relating
{\em entropies} for the three dichotomic random variables
$A$, $B$, and $C$.
We assume that one has access to the entropy of each variable
$H(A)$, $H(B)$, $H(C)$, as well as to the mutual entropy between each pair
of variables $H(A{\rm :}B)$, $H(A{\rm :}C)$, and $H(B{\rm :}C)$.
Here, the entropies are Shannon entropies~\cite{bib_shannon}, given by
\begin{equation}
H(A)= - \sum_a p(a) \log p(a)
\end{equation}
and the mutual entropies are defined by
\begin{equation}
H(A{\rm :}B) = H(A)+H(B)-H(AB)\;.
\end{equation}
The mutual entropy $H(A{\rm :}B)$ corresponds to the entropy shared by
$A$ and $B$, or in other words to the {\em information} about $A$ that
is conveyed by $B$ (or conversely). Physically, $H(A{\rm :}B)$ is
closely related to the correlation coefficient between $a$ and
$b$. To establish notation, let us also define the conditional
entropy $H(A|B)$ as the entropy of variable $A$ while ``knowing'', i.e.,
having measured, $B$,
\begin{equation}
H(A|B) = H(AB) - H(B)\;,
\end{equation}
which allows us to separate any entropy into a conditional and a mutual
piece with respect to another variable~\cite{bib_ash}:
\begin{equation}
H(A) = H(A|B) + H(A{\rm:}B)\;.
\end{equation}
For a three-variable system we can split {\em information} into
conditional and mutual information in the same fashion:
the information $H(A{\rm:}B)$, for example, can be split as
\begin{equation}
H(A{\rm:}B)=H(A{\rm:}B|C) + H(A{\rm:}B{\rm:}C)\;.
\end{equation}
Thus, a {\em conditional} information such as 
$H(A{\rm:}B|C)=H(AC)+H(BC)-H(C)-H(ABC)$ is that
piece of an information (between two variables) that is not shared by
a third variable, i.e., the information {\em conditional} on the third
variable. The piece of information
that {\em is} shared by a third variable can be written as
\begin{eqnarray}
\lefteqn{H(A{\rm :}B{\rm :}C)=H(A)+H(B)+H(C)} \nonumber \\
& & ~~~~~-H(AB)-H(AC)-H(BC)+H(ABC)\;.
\end{eqnarray}
Let us now construct Bell inequalities involving only
informations between pairs of variables (rather than correlation
coefficients). Relations between entropies are conveniently
represented by entropy Venn
diagrams~\cite{bib_meas,bib_reality}, and inequalities can easily be
read off them.
As shown in Fig.~\ref{fig_ternary}, the ternary entropy diagram for
the Bell variables $ABC$ has 
7 ($2^n-1$ with $n=3$) entries.  We use
the symbols $\alpha$, $\beta$, $\gamma$ 
for conditional entropies
[e.g., $\alpha=H(A|BC)$], $\bar\alpha$, $\bar\beta$, $\bar\gamma$
for conditional informations [e.g., $\bar\alpha=H(B{\rm:}C|A)$], and
denote by $\delta=H(A{\rm:}B{\rm:}C)$ the 
mutual information between the three Bell variables. Apart
from the marginal statistics of each of the variables $A$, $B$, and
$C$, experimentally we also have access to the the marginal statistics
of any pair ($AB$, $AC$, or $BC$), yielding six
constraints. Consequently, we do not have enough constraints to
completely fill in the entropy diagram of Fig.~\ref{fig_ternary}: the
missing constraint concerns the intrinsic three-body correlation which
is not fixed by two-body statistics.  The seven entries in the
ternary entropy diagram can thus be expressed as a function of the six
entropies $H(A)$, $H(B)$, $H(C)$, $H(A{\rm :}B)$, $H(A{\rm :}C)$,
$H(B{\rm :}C)$, plus a parameter $\delta$, the {\em inaccessible}
ternary mutual information.

\par

Despite this indeterminacy, the entries can be combined 
to give expressions {\em independent} 
of $\delta$, and which therefore can be expressed in terms of
measurable entropies only. More precisely, we find
\begin{eqnarray}
\alpha+\bar\alpha & = & H(A)+H(B{\rm :}C)-H(A{\rm :}B)-H(A{\rm :}C) 
\label{enteq1}\;,\\
\beta+\bar\beta & = & H(B)+H(A{\rm :}C)-H(A{\rm :}B)-H(B{\rm :}C) 
\label{enteq2}\;,\\
\gamma + \bar\gamma & = & H(C)+H(A{\rm :}B)-H(A{\rm :}C)-H(B{\rm :}C)
\label{enteq3}\;.
\end{eqnarray}
If $A$, $B$ and $C$ describe a classical system, it is known that
all the entries except $\delta$ are non-negative.
Indeed, monotonicity of Shannon entropies implies that conditional
entropies such as $\alpha=H(A|BC)$ are positive 
semi-definite~\cite{bib_wehrl}.
By the same token, conditional informations such as
$\bar\alpha = H(B{\rm :}C|A)$, as they describe information between
two variables when a third is known, are non-negative.
(This property is called strong 
subadditivity~\cite{bib_wehrl}.) The indeterminacy of $\delta$ can be
traced back to the freedom in the choice of a local hidden-variable
model to describe the marginal statistics, but its value is
unimportant as far as questions of locality are concerned.
From Eqs.~(\ref{enteq1}-\ref{enteq3}) it
follows straightforwardly that the three inequalities
\begin{eqnarray}
\label{eq_entbell1} H(A{\rm :}B)+H(A{\rm :}C)-H(B{\rm :}C) &\le& H(A) \\
\label{eq_entbell2} H(A{\rm :}B)-H(A{\rm :}C)+H(B{\rm :}C) &\le& H(B) \\
\label{eq_entbell3} -H(A{\rm :}B)+H(A{\rm :}C)+H(B{\rm :}C) &\le& H(C)
\end{eqnarray}
must be satisfied if the system $ABC$ is classical. These equations
therefore constitute entropic Bell inequalities. Note that
in the case where $A$, $B$, and $C$ have a uniform distribution, one has
$H(A)=H(B)=H(C)=1$;
the inequalities then become very similar to the standard ones 
[Eqs.~(\ref{eq_bell1}-\ref{eq_bell3})], but
relating  mutual entropies rather than correlation coefficients.
For instance, one can write
\begin{equation}
| H(A{\rm :}B)-H(A{\rm :}C)| + H(B{\rm :}C)\le 1 
\end{equation}
from Eqs. (\ref{eq_entbell2}) and (\ref{eq_entbell3}), in perfect
analogy with Eq.~(\ref{eq_bell}). More generally, the CHSH
inequalities for mutual entropies can be derived 
using the chain rule for entropies. The resulting inequality
\begin{equation}
H(A'{\rm:}B) + H(A{\rm:}C) - H(B{\rm:}C) + H(A{\rm:}A') \le 2\;,
\end{equation}
is similar in form to the traditional CHSH
inequality (see~\cite{bib_peres}), and 
implies the Braunstein-Caves inequality~\cite{bib_braunstein}
as well as Schumacher's quadrilateral inequality~\cite{bib_schumacher}.
\par

\begin{figure}
\caption{(a) Left-hand side $L_E$ of entropic Bell inequalities 
Eqs.~(\ref{eq_entbell1}-\ref{eq_entbell3})
for EPR Bell-measurements with $\theta=\pi/3.958$. The inequalities are
violated if $L_E>1$; (b)  Left-hand side $L_C$ of
conventional inequalities Eqs.~(\ref{eq_bell1}-\ref{eq_bell3}) at the
same angle.}
\vskip 0.25cm
\centerline{\psfig{figure=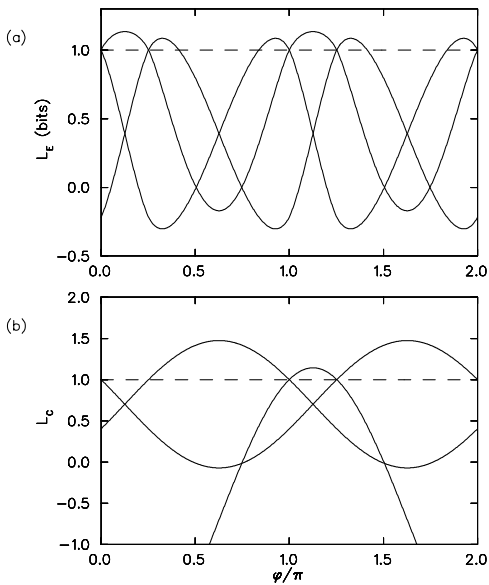,width=2.5in,angle=0}}
\label{fig_bellgraph}
\vskip -0.25cm
\end{figure}

The converse of the previous reasoning is most interesting. If the
data that are extracted from marginal statistics show that one
of the three entropic inequalities is violated, it implies that
one of the three inequalities $\alpha+\bar\alpha \ge 0$
(etc.) is violated. Therefore, since strong subadditivity of {\em quantum}
entropies~\cite{bib_wehrl} implies that $\bar\alpha$, $\bar\beta$, and
$\bar\gamma$ are always $\geq 0$,
one of the conditional entropies $\alpha$, $\beta$, or $\gamma$ {\em must} be
negative, which of course is classically forbidden.  Thus, a violation
of an entropic Bell inequality always goes hand in
hand with the appearance of a negative conditional entropy in
Fig.~\ref{fig_ternary}. This is the case for example in Bell
measurements of EPR pairs, as we show in more detail below. Therefore,
negative entropies automatically rule out a description of the system
in terms of local hidden variables (or an underlying
joint probability distribution). If there cannot be any such
description, it is well-known that the system in question is 
{\em non-separable}~\cite{bib_werner}. 
Equivalently, it is shown in Refs.~\cite{bib_neginfo,bib_meas} 
that the concavity
of conditional quantum entropies implies that any separable density
matrix is characterized by non-negative conditional entropies
(see also~\cite{bib_horo}).
In summary, the satisfaction of entropic Bell-inequalities, or
equivalently the non-negativity of the corresponding entropies, is a
{\em necessary} condition for separability, albeit not a sufficient one.
Let us show that this condition is distinct from the
one based on the satisfaction of traditional Bell inequalities by
considering as an example Bell experiments on EPR pairs. In this case,
because the outcomes $\pm 1$ occur with equal probability (1/2),
the correlation coefficient can be written as
$\langle ab\rangle  = 4 p_{++} -1 = 1 - 4 p_{+-}$, with $p_{++}$ ($p_{+-}$)
being the probability to observe aligned (anti-aligned) spins.
The mutual entropy (in bits) can then trivially be expressed in terms of the
corresponding correlation coefficient via
\begin{equation}
\hspace {-0.15 truecm} 
H(A{\rm:}B) = \frac12\log{(1-\langle ab\rangle^2)} + 
\frac{\langle ab\rangle}2\log{\left(\frac{1+\langle ab\rangle}
{1-\langle ab\rangle}\right)} \;.
\end{equation}
Using the standard quantum results for the correlation
coefficients, i.e., $\langle ab\rangle = -\langle a'b\rangle=
\cos(\theta)$, $\langle ac\rangle = -\cos(\phi)$, and $\langle
bc\rangle = -\cos(\theta-\phi)$, we plot in Fig.~\ref{fig_bellgraph}a
the left-hand side of Eqs.~(\ref{eq_entbell1}-\ref{eq_entbell3}) as a
function of $\phi$ for the ``most violating'' angle $\theta =
\pi/3.958$ (the maximum violation occurs at $\phi=\theta/2$).  Note
that the conventional Bell inequalities
Eqs.~(\ref{eq_bell1}-\ref{eq_bell3}) are maximally violated at
a different angle $\theta = \pi/3$. Nevertheless, we
have plotted the left-hand side of these equations at the same angle
$\theta$ as the entropic ones for comparison in
Fig.~\ref{fig_bellgraph}b. Despite the similarity in the structure of
the equations, the violation of one conventional Bell inequality does
not necessarily imply the violation of an entropic one, or vice versa.

\par 

We have derived entropic Bell inequalities by demanding that
the conditional entropies arising in
the ternary entropy diagram for Bell variables be
non-negative, providing a necessary condition for separability.
The experimental violation of Bell inequalities,
traditionally interpreted as ruling out the existence of a joint
probability $p_{abc}$, therefore also reflects the appearance of
negative conditional entropies in Bell-type measurements. 
In fact, these experiments do not rule out a description
in terms of an underlying joint {\em density matrix} $\rho_{ABC}$.
Yet, the latter does not describe {\em three} physical
systems as the EPR experiment only involves {\em two} detectors. 
Because of the degree of freedom involved with the choice of
$\delta$, such a $\rho_{ABC}$ cannot be
constructed explicitly. We are therefore uncertain as to the physical
interpretation of $\rho_{ABC}$,
a difficulty inherent to {\em independent} Bell-type measurements 
on identically prepared systems. It has recently been 
suggested that
{\em consecutive} measurements performed on a single quantum system
are more apt at revealing ``hidden nonlocality''~\cite{bib_popescu}.
It might therefore prove to be fruitful to apply the present analysis
to such situations.

This work was supported in part by NSF Grant
Nos. PHY 94-12818 and PHY 94-20470.

\end{multicols}

\begin{references}
\bibitem{fn1}One of the orientations ($z$-axis) is the same
for both observers; if this constraint is relaxed, one obtains the
generalized CHSH inequality~\cite{bib_CHSH}. For simplicity, we 
confine our attention to the standard Bell inequality here.
\end{references}
\end{document}